# Dynamic IoT Choreographies

## Managing Discovery, Distribution, Failure and Reconfiguration

**Jan Seeger**
Siemens AG / TU Munich

**Rohit A. Deshmukh**
TU Darmstadt

**Vasil Sarafov**
TU Munich

**Arne Bröring**
Siemens AG

The Internet of Things is growing at a dramatic rate and extending into various application domains. We have designed, implemented and evaluated a resilient and decentralized system to enable dynamic IoT choreographies. We applied it to maintaining the functionality of building automation systems, so that new devices can appear and vanish on-the-fly.

Pervasive connectivity in machine to machine communication as well as advancements in sensor and actuator technology has given rise to the Internet of Things (IoT). For companies such as Siemens, the IoT concept plays a tremendous role for their ongoing evolution towards full digitalization. Smarter cities, eHealth systems, or Industry 4.0-enabled manufacturing plants are transforming into IoT environments. In this paper, we take an application from the building automation domain as an example to illustrate our solution. While building automation systems (BAS) have certain specific constraints and characteristics, the solutions presented in this paper are applicable to other IoT domains.

In today's building infrastructures, we find heterogeneous devices such as lights, switches, window shutters, air conditioners, light sensors, or thermostats, which are being IoT-enabled. I.e., the communication with such components is based on Internet and Web technologies, such as the HTTP or CoAP, REST interfaces and data are exchanged via JSON. In the future, such IoT environments will become very dynamic. New devices will vanish and appear on-the-fly during their lifetime or change their properties.

As an example, imagine a reconfigurable multi-purpose room that users can partition into multiple rooms by movable walls. When users move walls in the room, the installed equipment should automatically adapt to the new room configuration, which requires dynamic reconfiguration of the building system.

The operation of such a dynamic room today requires the involvement of technicians at each reconfiguration stage, or provide a suboptimal user experience, with light switches not operating the correct lights, or illuminating only a portion of the room. For reasons of brevity, we have confined our example in this paper to a simple switch-light system which might appear in such a dynamic meeting room, where a switch operates an indeterminate number of lights that may change dynamically.

For these reasons, we have chosen to focus on the domain of building automation systems for our application example. BAS have additional constraints that are not present in general IoT systems, but the solutions presented in this paper are applicable to other domains.

We approach these challenges by extending semantic application descriptions (called *recipes*) with constraints to enable dynamic and automatic reconfiguration of applications. These recipes are executed as a distributed and autonomous *choreography* that behaves according to constrains configured during design time. To ensure reliability, we provide a novel mechanism based on accrual failure detection that detects failures in the distributed system and recovers dynamically using the defined constraint system. Finally, we evaluate our implementation of the approach through an application demonstration. Further, we test the performance of the failure detection mechanism and compare it to two related approaches. This work builds up on our previous works on semantically-enabled IoT device composition [1], [2] and marks a further milestone on our continued research path.

## RELATED WORK

When new devices are added to an IoT environment such as a building automation system (BAS), they need to be connected physically, and the software on the controller need to be reparametrized and reconfigured. Today, this task requires substantial effort and profound expertise. Therefore, it is not possible to enable dynamic behavior in such a traditional BAS.

Thuluva et al. [3] employ Semantic Web technologies to enable low-effort engineering of automation systems. However, in their Industry 4.0 approach, they do not focus on the runtime aspects and dynamic reconfiguration or failure detection yet. Using our integration of failure detection and rule-based interaction, the system can be dynamically reconfigured at runtime without user interaction.

The "scope" concept described by Mottola et al. in [4] is similar to our concept of offering selection rules. However, the system does not support dynamic scopes, whereas our system supports the modification of device descriptions on-the-fly during the operation of the system and thus the dynamic reconfiguration of devices.

The DS$^2$OS system by Pahl et al. [5] supports dynamic addition and removal of services through the use of the blackboard pattern for communication, but no implicit mechanism is provided to deal with newly appearing devices, as this must be implemented manually by the user. Our system does not require special handling of dynamic behavior, only a specification of the requirements of the service.

Our system composes different services offered by web-enabled IoT offerings to form an application. Web service composition can be classified into two types, service *orchestration* and service *choreography*, based on the way the participant services interact (for more information, see Sheng et al.[6]). Applications in the web-services composition field follow the orchestration approach, and do not cover reliability facets. Khan et al. [7] propose a reliable IoT infrastructure, but focus solely on communication links. They do not examine the aggregation of nodes into larger applications as provided by our recipe model.

A basis for reliability in IoT environments is the detection of failures of involved devices and services. A high level approach for such failure detection for living spaces with emphasis on system maintenance is presented in [8]. Kodeswaran et al. suggest detecting activities of daily living in IoT home environments which are used to reason about the expected future degradation of the connected devices. This approach is valid only for smart home applications and is not suitable for explicit failure detection in general purpose IoT systems. The failure detector described in this work is not restricted to any specific application areas.

In [9], Chetan et al. present the Gaia middleware for pervasive distributed systems which implements a fault-tolerant mechanism. However, the failure detection in this system is centralized, which means that the controller is required to be available during the operation of the services. In our recipe system, this is not the case, as nodes can detect the failure of other nodes without communicating with the controller.

In [10], De Moraes Rosetto et al. present the design of an unreliable failure detector for ubiquitous environments, which supports grouping and assigning of different impact factors of nodes. A distributed fault detection algorithm in a sensor application based on trust management is presented by Guclu et al. in [11]. They propose to use a method that relies on evaluating statements about trustworthiness of aggregated data from neighbors. This approach is valid only for homogeneous networks where data of the same kind is collected. Thus, it is not applicable in our recipe system, where we use a more general failure detector.

# DYNAMIC DISCOVERY, DISTRIBUTION AND CONFIGURATION

Today, IoT devices are typically composed by a central entity as a service *orchestration*. The central entity, which providers typically host in the cloud, schedules and calls different device and service functions. Such an orchestration entails a single point of failure, as well as latency, network and privacy disadvantages. Our system instead forms the components into a service *choreography*, where, during operation of the service, devices speak to each other directly, without going through a third party. This takes advantage of the growing "smartness" of automation devices and aims to partly mitigate the dependency on a single centralized orchestration point during operation of the service. Only during the reconfiguration, the system still requires a central controller. We describe a possible solution to this problem in the "Conclusions" section.

At the heart of our system, we use the *recipe* concept to represent the abstract structure of IoT device interactions [1]. A recipe describes the dataflow between devices through *ingredients* and *interactions*, as shown in Figure 1. Ingredients represent a class of devices or services through a semantic category and a number of inputs and outputs that carry data type information. The category describes the *kind* of thing that this ingredient represents. Using semantic concepts, the category description can be made as fine- or coarse-grained as desired. The inputs and outputs describe the type of data that this ingredient requires for operation, and the results of its operation. Interactions describe the data flow between offering inputs and outputs, and must take place between offering inputs and outputs with matching types. When an offering receives a set of complete inputs, a computation or measurement is executed. The result of this computation is sent to the outputs, and along the interaction edge to the next offering's inputs.

This model is also suitable to express "part-of" relationships for devices consisting of several ingredients. To model a lamp with color temperature and brightness inputs, the device would be modeled as two separate ingredients that have a "device" non-functional property. In the recipe, the author would then constrain the ingredients to be part of the same device via OSRs.

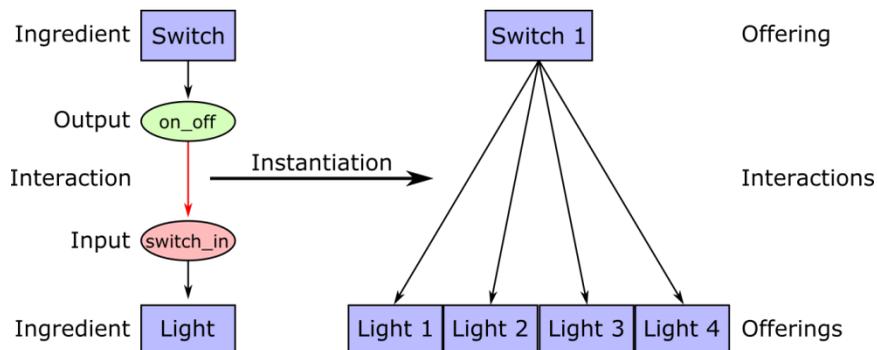

Figure 1: Instantiating a recipe of ingredients to running choreography of offerings.

Recall that a recipe only represents a *template* for a system. Ingredients represent placeholders for concrete devices or services. To operate, a recipe needs to be instantiated, which means that these placeholders need to be replaced with actual services or devices. We call these services or devices *offerings* and the process of replacing ingredients with offerings *instantiating* a recipe.

Offerings are described similarly to ingredients (category, inputs and outputs), but correspond to concrete device functionality. Additionally, offerings contain non-functional information about their current state (such as location or administrative information), as opposed to functional information (inputs, outputs, category and implementation). *Interactions* describe the data flow between offering inputs and outputs, and must take place between offering inputs and outputs with matching types. When an offering receives a set of complete inputs, a computation or measurement is executed. The result of this computation is sent to the outputs, and along the interaction edge to the next offering's inputs.

To extend the applicability of the recipe concept to other domains, we have introduced two new concepts into the system, *offering selection rules* and *recipe runtime configurations*.

*Offering selection rules* (OSRs) describe additional requirements on an offering's non-functional properties that must be met in order for the offering to be considered for an ingredient's replacement [2]. All non-functional properties of offerings can be restricted, and restrictions can be combined with Boolean operators. Also, cardinality restrictions allow limiting the number of offerings that replace an ingredient between a lower and upper bound. Offering selection rules allow the expression of more complex system requirements than through semantic matching on functional properties alone.

*Recipe runtime configurations* (RRCs) contain information about a specific instance of a recipe. Recipes may be instantiated multiple times, possibly with different OSRs, and each instantiation forms a new RRC. The process of selecting offerings that fulfill the requirements to be part of a recipe is called *offering discovery*. In this process, the matching of category as well as input and output types of an ingredient in the RRC is determined. This process strongly relies on the semantic annotation of the descriptions of offerings and recipes. The recipe runtime configuration allows the instantiation of a single recipe multiple times with varying OSRs. Currently, this is done through repeated instantiation of predefined static templates, and RRCs replace this mechanism with a more flexible approach.

Using these concepts, we have built a system that enables (1) the specification of recipes in a graphical recipe editor, (2) browsing and viewing of recipes, (3) creating and modifying RRCs, and (4) building executable choreographies from RRCs. The relevant components are: The *controller*, which is hosts the graphical interface (described in more detail in Thuluva et al. [1]) and enables the creation of choreographies from recipes, a semantic *database* for persistence and semantic operations and finally the *engine*. These components and their interaction are shown in Figure 2.

The engine is a piece of software hosted on smart devices that allows the running of choreographies. Each engine carries a description of the offering it provides (the so-called offering description) and provides an interface to realize the operation of choreographies. The engine uses the concepts described by [12] to adapt heterogeneous devices to the standardized recipe interface.

Controller and engine work together to enable the running of distributed choreographies generated from RRCs. When an engine is added to the network or configuration of the engine changes, it registers at the controller with its offering description. The controller then runs the offering discovery process, finding all RRCs that the newly added engine should be part of. From the recipe corresponding to the RRC, the controller derives all offerings that the engine should communicate with. This communication information is encoded into a so-called interaction descriptor (InDes) for each engine part of the RRC and distributed.

An interaction descriptor contains information about the components communication behavior derived from the recipe. It contains information on which inputs should be mapped to which outputs, and information on the failure detector configuration to use for this communication link.

After receiving an interaction descriptor, the engine can fulfill service execution and failure detection autonomously, without communication with the controller.

Using the mechanism described above, a choreography is created and run. The engine receives input data (sent via HTTP PUT requests encoded as JSON data) and transfers that data to the implementing service or device. This communication is shown by black arrows in Figure 2. The service creates new output that the engine again encodes as JSON and sends to the REST endpoints described in the output section of its interaction descriptor. Once the choreography is created, the controller is no longer required, and the functionality described by the recipe is provided by offerings without centralized coordination.

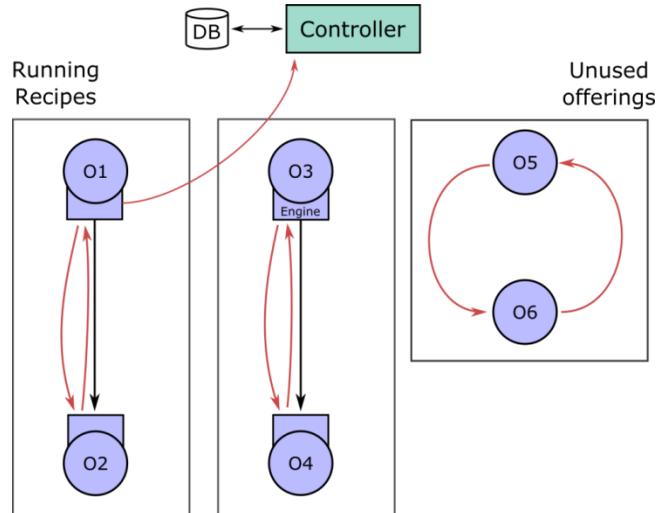

Figure 2: Failure detection in a distributed IoT environment. Red arrows indicate failure detection communication, black lines indicate application communication.

An example of this approach for a light control recipe is shown in Figure 2: The recipe consists of two switches (*O1* and *O3*) controlling lights (*O2* and *O4*). Two offerings are currently not part of a recipe (*O5, O6*), and thus do not exchange recipe data. Each switch controls a single light. The straight black arrows show recipe data flow, while the red arrows show monitoring links created from this recipe data flow. It can be seen that all recipe data flows result in exactly one monitoring link being created with additional backlinks to initial nodes of the recipe. Unused offerings that are not part of a recipe are connected into a ring by monitoring links to maintain readiness. When failure is detected by a node according to the recipe-specific failure detection parameters, a notification is sent to the controller (represented by the red arrow to the controller) and the controller reruns the offering selection mechanism to replace the failed offering.

# FAILURE DETECTION

As outlined above, choreographies haven clear advantages over centralized compositions. However, when minimizing the role of a centralized component, detection of component failures becomes more challenging. In the case of running choreographies for automation systems or the IoT, a mechanism for failure detection is especially important, as failures have a more severe impact and are more likely compared to traditional web services.

Hence, our aim is to ensure that the detector notices the failure of a choreography, and the controller recovers the failed choreography, so that our system can provide services with minimal downtime.

We have designed a failure detection algorithm (FD) that takes these factors into account and provides the engine with high-level information on the reliability of nodes it communicates with. The algorithm is self-adaptive and flexible and thus can be applied not only in building automation systems, but also in other IoT applications. Additionally, the failure detector uses constant memory over the full range of parameters, which is important for embedded and constrained devices. The failure detector is based on the concept of accrual failure detection described by Hayashibara et al. [13]. We have named our approach *iota-FD* ("Internet of Things Accrual Failure Detector").

The failure detection algorithm is deployed on each node in a system. A client is a node that is monitored by the algorithm. A server is a node that monitors clients. If needed, a node can be both a client and a server.

An accrual function as defined in [13] is a suspicion function whose output value represents the confidence of the failure detector that a monitored device has crashed. Values close to 0 imply that the device is working correctly, while greater values indicate higher confidence that a monitored device has crashed. In our system, we use a per-service suspicion threshold, above which we consider a device crashed. When suspicion approaches this value, our system can take mitigating action such as migrating data or find a suitable replacement for a service.

We implement the accrual suspicion function using the Chebyshev one-sided inequality to easily approximate the probability of the presence of a crashed device. Using a recursive computation about the mean value and variance of inter-arrival heartbeat times, we can efficiently calculate and update the suspicion level over time while using constant memory. This is an advantage to implementations using the empirical distribution function, which requires storage of the complete list of timestamps.

In addition, iota-FD provides measurements about the critical resources left in a monitored device (e.g. battery level). With the help of the collected measurements, the FD builds a Lagrange polynomial, which estimates the depletion/augmentation of the resources. This information can be used to indicate when devices need to be serviced.

Finally, iota-FD measures the quality of the end-to-end communication link between a server and a client. This is achieved by estimating the current packet drop rate. Using a weighted exponential moving average, the calculated drop rate changes over time and does not remain static, i.e., the iota-FD is able to "forget" past disruptions in the network and "learn" new, once they have occurred. This information can be used to differentiate between crashed devices and devices whose heartbeats are lost because of bad network conditions.

Summarizing the above, the suspicion function *s(t)*, the packet loss predictor *p(t)* and the resource predictor *r(t)* are shown in the following definition:

*Definition*: With the last *n* heartbeats received at times $t_0 < t_1 < t_2 \ldots < t_n$ and the current timestamp iota-FD computes:

- $s(t) = -log_{10}(Pr[X > t - t_{n-1}]) = -log_{10}\left(\frac{\sigma(t)^2}{\sigma(t)^2 + (t - t_{n-1} - \mu(t))^2}\right)$, where *X* is the time delay until a new heartbeat will be received in the future and μ and σ are respectively the mean and variance of the distribution of heartbeat inter-arrival times.

  $\mu(t)$ is calculated from the current sum of the inter-arrival times of timestamps ρ divided by the number of received timestamps *n*, while $\sigma(t)^2$ is calculated from the current sum of the squares of the inter-arrival times of all timestamps κ and ρ using the equality $\sigma(t)^2 = \left(\frac{\kappa}{n} - \left(\frac{\rho}{n}\right)^2\right)\left(\frac{n}{n-1}\right)$.

  Both ρ and κ are reset when $\omega_{max}$ timestamps have been received. To reduce the impact of this reset on the quality of the period estimation, old estimations are used until a certain minimum number of timestamps $\omega_{min}$ has been received.

  Thus, the calculation of the suspicion function requires enough storage for two unsigned integers large enough to not overflow within the chosen learning window, and one integer to store the current number of timestamps. The memory required for failure detection is thus independent of the learning window, an important advantage over competing implementations.

- $p(t) = p(t_n) = \alpha \frac{\text{lastburstlength}}{\text{totalslidingwindowlength}} + (1-\alpha)p(t_{n-1})$, where $\alpha$ is a parameter that determines the speed of learning (first term) and forgetting (second term) of future and past bursts respectively.

- $r(t) = \sum_{j=0}^{3}\left[\rho_{t_{n-j-1}} \prod_{i=0, i \neq j}^{3} \left(\frac{t-t_{n-i-1}}{t_{n-j-1}-t_{n-i-1}}\right)\right]$ where $\rho \in [0,1]$ is received with every heartbeat and expresses the level of resources currently left at a client device. No assumptions are made about how $\rho$ is calculated by the client.

Iota-FD provides a wide variety of parameters to configure failure detection on a per-node and per-service basis: The learning window size used for the estimation of heartbeat timings of the algorithm can be adjusted freely. The estimator for heartbeat timings is self-adapting, allowing clients to adjust their heartbeat frequency dynamically based on external factors such as battery charge without requiring explicit configuration. Alternatively, the heartbeat period can be adjusted based on the "importance" of the service in the recipe or other factors depending on the recipe definition.

Additionally, the information computed through *p(t)* and *r(t)* allow the definition of fine-grained policies for the application (such as "if suspicion is medium-high and no packet loss bursts were registered, migrate service data in preparation for a device crash" or "if the resources will be below a certain threshold in 2 hours, notify the administration to schedule a servicing routine"). Service definitions can include such policies to define QoS requirements on the service, and continually evaluate the current quality of the service. We plan an implementation of this functionality in the future.

# EVALUATION

To measure the behavior of our failure detector as compared to other state-of-the-art failure detectors suitable for usage in distributed IoT systems, we have measured the behavior of our approach compared with two other failure detectors:

First, we compare iota-FD against the "Phi-Accrual" failure detector defined by Hayashibara et al. [13], which uses a different implementation of the suspicion function compared to iota-FD. It is used in the Apache Cassandra distributed database and Akka distributed programming framework. The Phi-Accrual detector always assumes that the inter-arrival times of the heartbeats are normally distributed. It calculates the above-mentioned probability using the same estimators but applies them to the normal cumulative distribution function instead of the Chebyshev inequality as iota-FD does.

Second, we evaluate the "Adaptive" failure detector proposed by Satzger et.al. in [14] that uses a different definition of the suspicion function, where *s(t)* ∈ [0, 1] is the probability that a node has crashed. The value of *s(t)* is computed using the empirical distribution function applied to a list of stored heartbeat inter-arrival times.

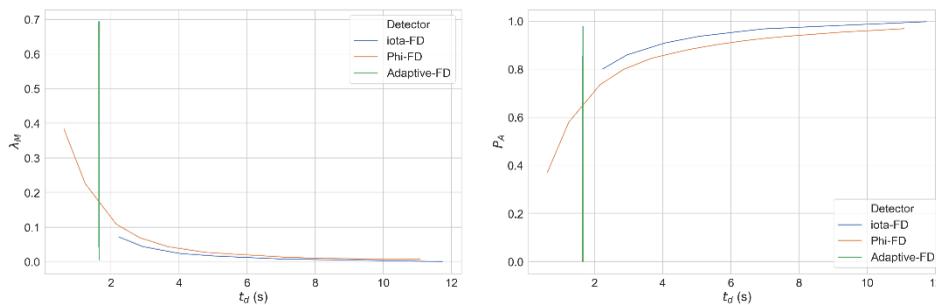

Figure 3: Mistake rate vs. detection time (left), query accuracy probability vs. detection time (right)

As the related work focuses on pure accrual failure detection, we will not benchmark the resource and packet drop estimators but focus on the accrual functions. In this evaluation, we will evaluate the tradeoffs between one performance (detection time) and two accuracy QoS metrics (average mistake rate and query accuracy probability) defined by Chen et al. [15] for the three failure detectors with comparable parameterizations. Detection time is the time a failure detector needs to permanently start suspecting a crashed device, average mistake rate measures how often a correct device is wrongly suspected to have crashed and query accuracy probability measures the probability that when queried at a random time, a failure detector will answer correctly whether a device is faulty or correct.

These parameters are the threshold $U$, the learning window size $\omega_{max}$, the refresh duration $\omega_{min}$ (only iota-FD), and the smoothness parameter $\alpha$ (only Adaptive).

The learning window size $\omega_{max}$ was fixed to 500 for all failure detectors as a configuration that is suitable for the types of constrained devices that are targeted. The rest of the parameters were varied to generate the QoS graphs in Figure 3.

The simulations were run as follows: Heartbeat times were generated using a normal distribution with a mean of 1 and a variance of 9. This distribution stems from application and network level variance as to the generation of heartbeats. It models a case where the majority of heartbeats are received with delays between zero and five time units, most of which are located in the surroundings of one time unit. Additionally, the burst packet loss model from ns-3[1] was used to model packet loss in a wide area network. The failure detector algorithms were applied to the resulting heartbeat times, and the values of the suspicion function were sampled. From this data, mistake rate, detection time and query accuracy probability were computed for different thresholds $U$. The values for corresponding $U$ are joined to illustrate the tradeoffs between accuracy and performance metrics.

As it is shown in Figure 3, the Adaptive failure detector is a special case: Its suspicion function is a discrete function, and the learning buffer was not large enough to generate a smooth curve. A small learning window of 500 heartbeats was chosen as IoT devices are constrained. This resulted in a very low detection time of the Adaptive failure detector (lower than both other failure detectors), but also a correspondingly high mistake rate can be observed. Additionally, it is not possible to adjust the tradeoff between both parameters by varying $U$. On the other hand, both the Phi and iota-FD approaches allow this tradeoff, with the iota-FD performing better in both tradeoffs, which results in a lower detection time for a certain mistake rate or query accuracy probability and vice versa.

In addition to the above evaluation of the failure detection mechanism, we have built a small application demonstration using two "switch" offerings (Raspberry Pi single board computers with attached pushbuttons) and two "light" offerings (one smart office light accessible via CoAP, and one Philips Hue[2] light accessible via HTTP REST). We created a recipe that allows control of offerings of the category "Light" located in "Room A" from all offerings with the category "Switch" in "Room A". We also added a "maximum cardinality of 1" constraint to the switch ingredient. The two switch offerings were connected to the network, and one switch was selected for control of the lights. To see the reliability functionality in action, we then removed power to the switch currently controlling the light and observed failover to the second switch within 15 seconds. A recording of this application example's operation accompanied with narration is available at https://mediatum.ub.tum.de/1403134?show_id=1432951.

## CONCLUSIONS

This work proposes an architecture that allows dynamic reconfiguration of IoT choreographies based on the iota-FD mechanism for failure detection. We applied our approach in a building automation scenario and evaluated the iota-FD mechanism in comparison to two other approaches. We found the recipe system represents a suitable programming approach for dynamic automation systems and the iota-FD is well suited for IoT use cases where small learning buffers are assumed due to constrained devices. Iota-FD also provides a wide array of parameters that can be adjusted on a per-application basis. Using recipes, dynamic choreographies can be created that self-adapt to changing device states without user interaction.

---

[1] Described at https://www.nsnam.org/doxygen/classns3_1_1_burst_error_model.html#details

[2] https://www2.meethue.com/en-us

Having implemented and evaluated our reliability approach for managing dynamic IoT choreographies, we are planning to extend both the evaluation and implementation. On the implementation level, we plan to take advantage of software defined networking (SDN) technology to be able to replace offerings in a network-aware fashion. Additionally, we will evaluate the implementation of a distributed controller using replication and leader selection.

Beyond reliability, we will extend the capabilities of the offering description to not only describing choreographies of web services, but allowing *computation* offerings as well, with features comparable to fog computing technology, i.e., being able to deploy computation offerings to a local "cloud" and being able to seamlessly migrate such computations between nodes. Further, an important aspect of future work will be the validation of a reconfigured choreography after a recovered failure. Here, this challenge can be either tackled in advance of execution, by introducing further semantics to device descriptions, or at runtime, by performing consistency checks after mediation.

# REFERENCES

[1] A. S. Thuluva, A. Bröring, G. P. Medagoda, H. Don, D. Anicic, and J. Seeger, "Recipes for IoT Applications," in *Proceedings of the Seventh International Conference on the Internet of Things*, New York, NY, USA, 2017, pp. 10:1–10:8.

[2] J. Seeger, R. A. Deshmukh, and A. Bröring, "Running Distributed and Dynamic IoT Choreographies," in *2018 IEEE Global Internet of Things Summit (GIoTS) Proceedings*, Bilbao, Spain, 2018, vol. 2, pp. 33–38.

[3] A. S. Thuluva, K. Dorofeev, M. Wenger, D. Anicic, and S. Rudolph, "Semantic-Based Approach for Low-Effort Engineering of Automation Systems," in *On the Move to Meaningful Internet Systems. OTM 2017 Conferences*, 2017, pp. 497–512.

[4] L. Mottola, A. Pathak, A. Bakshi, V. K. Prasanna, and G. P. Picco, "Enabling Scope-Based Interactions in Sensor Network Macroprogramming," in *2007 IEEE International Conference on Mobile Adhoc and Sensor Systems*, 2007, pp. 1–9.

[5] M.-O. Pahl, "Data-centric service-oriented management of things.," in *IFIP/IEEE International Symposium on Integrated Network Management, IM 2015, Ottawa, ON, Canada, 11-15 May, 2015*, 2015, pp. 484–490.

[6] Q. Z. Sheng, X. Qiao, A. V. Vasilakos, C. Szabo, S. Bourne, and X. Xu, "Web services composition: A decade's overview," *Inf. Sci.*, vol. 280, pp. 218–238, Oct. 2014.

[7] W. Z. Khan, M. Y. Aalsalem, M. K. Khan, M. S. Hossain, and M. Atiquzzaman, "A reliable Internet of Things based architecture for oil and gas industry," in *2017 19th International Conference on Advanced Communication Technology (ICACT)*, 2017, pp. 705–710.

[8] P. A. Kodeswaran, R. Kokku, S. Sen, and M. Srivatsa, "Idea: A System for Efficient Failure Management in Smart IoT Environments," in *Proceedings of the 14th Annual International Conference on Mobile Systems, Applications, and Services*, New York, NY, USA, 2016, pp. 43–56.

[9] S. Chetan, A. Ranganathan, and R. Campbell, "Towards fault tolerance pervasive computing," *IEEE Technol. Soc. Mag.*, vol. 24, no. 1, pp. 38–44, Spring 2005.

[10] A. G. De Moraes Rossetto *et al.*, "A new unreliable failure detector for self-healing in ubiquitous environments," in *Proceedings - International Conference on Advanced Information Networking and Applications, AINA*, 2015.

[11] S. O. Guclu, T. Ozcelebi, and J. Lukkien, "Distributed Fault Detection in Smart Spaces Based on Trust Management," *Procedia Comput. Sci.*, vol. 83, pp. 66–73, Jan. 2016.

[12] A. Bröring *et al.*, "Enabling IoT Ecosystems through Platform Interoperability," *IEEE Softw.*, vol. 34, no. 1, pp. 54–61, Jan. 2017.

[13] X. Défago, N. Hayashibara, R. Yared, and T. Katayama, "The Φ Accrual Failure Detector," in *Reliable Distributed Systems, IEEE Symposium on(SRDS)*, 2004, pp. 66–78.

[14] B. Satzger, A. Pietzowski, W. Trumler, and T. Ungerer, "A New Adaptive Accrual Failure Detector for Dependable Distributed Systems," in *Proceedings of the 2007 ACM Symposium on Applied Computing*, New York, NY, USA, 2007, pp. 551–555.

[15] W. Chen, S. Toueg, and M. K. Aguilera, "On the quality of service of failure detectors," *IEEE Trans. Comput.*, vol. 51, no. 1, pp. 13–32, Jan. 2002.

## ABOUT THE AUTHORS

**Jan Seeger** is a PhD researcher at Siemens AG as well as at the Technical University of Munich. He is active in the areas of Internet of Things and automation research, and how semantic technologies can improve the engineering of automation systems. He holds a M.Sc. in computer science from the TU Munich. He can be reached at jan.seeger@siemens.com.

**Rohit A. Deshmukh** holds a M.Sc. in Distributed Software Systems from the Technische Universität Darmstadt and has been with Siemens' corporate research unit in Munich while contributing to this work. His research interests include distributed software systems, the Internet of Things, peer-to-peer systems and the Semantic Web.

**Vasil Sarafov** is a M.Sc. student in computer science at TU Munich. His interests include distributed and embedded systems, algorithms and data structures. He can be reached at sarafov@cs.tum.edu.

**Arne Bröring** is a senior researcher at Siemens' corporate research unit in Munich and the technical coordinator of the BIG IoT project. His research interests include the Internet of Things, Sensor Web, and the Semantic Web. He received a Ph.D. in geoinformatics from the University of Twente (NL).